\documentclass[aps,prb,twocolumn,groupedaddress,showpacs]{revtex4}

\usepackage{graphicx}
\usepackage{dcolumn}
\usepackage{bm}
\usepackage{amsmath,amssymb}
\usepackage{color}
\usepackage [latin1]{inputenc}

\begin{document}


\title{Spin excitation of the Heisenberg antiferromagnet with frustration: 
from the bounce-lattice antiferromagnet 
through the maple-leaf-lattice antiferromagnet 
to the exact-dimer system
}



\author{Hiroki Nakano$^1$ and T\^oru Sakai$^{1,2}$}
\affiliation{$^1${Graduate School of Science, University of Hyogo, Kouto 3-2-1, Kamigori, Ako-gun, Hyogo 678-1297 Japan}\\
$^2${National Institute for Quantum Science and Technology (QST) SPring-8, Kouto 1-1-1, Sayo, Sayo-gun, Hyogo 679-5148 Japan}
}


\date{\today}

\begin{abstract}
The spin-$S$ Heisenberg antiferromagnet on the two-dimensional lattice
is investigated for $S=1/2$ and $S=1$. 
We consider interaction at isolated dimers ($J_{\rm d}$) 
and interaction bonds that form the bounce lattice ($J_{\rm b}$).  
For $J_{\rm d}=J_{\rm b}$, the system is reduced
to the maple-leaf-lattice antiferromagnet. 
We primarily conduct highly parallelized numerical diagonalization 
to examine the spin excitation gap above the ground state 
for various $J_{\rm b}/J_{\rm d}$ cases.  
For $S=1/2$, we report calculations for a 42-site cluster 
that has not been previously treated.  
The $S=1$ case is examined for the first time for clusters up to 24 sites. 
Regardless of whether $S=1/2$  or 1, 
we find that the system has a gapped nature 
for small $J_{\rm d}/J_{\rm b}$ 
and becomes gapless at $J_{\rm d}/J_{\rm b}\sim 1.4$. 
For $S=1$, we also find that another gapped region appears 
between the gapless case at $J_{\rm d}/J_{\rm b}\sim 1.4$ 
and the boundary of the exact-dimer phase. 
\end{abstract}


\maketitle


\section{Introduction}

Frustration in magnetic materials is widely considered 
an important source of exotic quantum states. 
As typical cases for the frustrated magnets, 
antiferromagnets on the kagome and triangular lattices 
have been intensively and extensively investigated 
both experimentally and theoretically. 
One index for characterizing the lattice structure 
is the coordination number $z$, which is 
the number of bonds linked to each spin site: 
$z=4$ for the kagome lattice 
and 
$z=6$ for the triangular lattice. 
In general, magnets with a smaller $z$ are considered 
to have strong quantum effects. 

The maple-leaf-lattice antiferromagnet ($z=5$) 
falls between these two cases 
and 
has been studied\cite{Betts,
Schulenberg_Richter_APPA2000,Shumalfuss_Richter_PRB2002,
Farnell_Richter_PRB2011,Farnell_Richter_PRB2014,Farnell_PRB2018,
Ghosh_PRBL2022,Ghosh_PRB2023,Gresista_PRB2023,Ghosh_JPCM2024,
Gembe_PRB2024,Ghosh_PRB2024,Beck_PRB2024,Maldonado_Mg3TeO6_PRB2025,
Ghosh_PRB2025,Hutak_PRB2025,Gresista_arXiv2025}. 
In some of these references, Richter played an important role. 
One major characteristic is that 
the ground state is the exact-dimer state 
when the dimer interaction is sufficiently large, 
which is shared by the Heisenberg antiferromagnet 
on the Shastry-Sutherland lattice\cite{ShastrySutherland1981}. 
The discovery of a good candidate material SrCu$_2$(BO$_3$)$_2$ 
for realizing the $S=1/2$ Shastry-Sutherland 
antiferromagnet\cite{Kageyama_PRL1999} 
has strongly prompted further investigations. 
For cases of the spin larger than $S=1/2$, 
the specific behaviors had not been well understood 
until numerical results were reported concerning finite-size clusters 
for $S=1$ and $S=3/2$ of this model\cite{HNakano_TSakai_JPCM2024}. 
Consequently, 
the Shastry-Sutherland model is characterized by 
the exact-dimer phase where the interaction 
at dimer pairs is strong,  
the N$\acute{\rm e}$el-ordered phase where the interaction is weak, and
furthermore an intermediate region between these two phase 
regardless of $S$. 

For the maple-leaf lattice, however, 
when the interaction at dimer pairs vanishes, 
the system is reduced to the bounce lattice. 
The antiferromagnet of the cases from the bounce lattice 
through the maple-leaf lattice to the isolated dimers 
was examined\cite{Farnell_Richter_PRB2011}. 
This reference found that the exact-dimer phase exists 
where the interaction at dimer pairs is sufficiently strong.  
However, triangular structures remain locally 
in the case of the bounce-lattice antiferromagnet.  
Thus, even when the interaction at dimer pairs vanishes, 
frustration does not completely disappear. 
Recent theoretical efforts have continued to explore these systems and 
generalized cases involving more extensive factors from various perspectives.
Even in the studies based on the tensor-network calculations, 
different conclusions concerning whether the long-range order is 
present or absent in the ground state of the bounce-lattice antiferromagnet  
were derived; Ref.~17 
presented 
the non-magnetic character on one hand while 
Ref.~19 
on the other hand showed 
small but nonzero long-range order. 
Experiments on the $S=1/2$ case was conducted and 
Cu$_6$IO$_3$(OH)$_{10}$Cl was reported 
to be a candidate material\cite{Haraguchi_Cu6IO3OH10CL_PRB2021}.
A tight-binding model of electrons on the maple-leaf lattice is 
also studied\cite{He_PRB2024}.

The present study has two main objectives. 
The first is to use numerical diagonalization to 
reexamine the $S=1/2$ case for the Heisenberg antiferromagnet 
from the bounce lattice through the maple-leaf lattice 
to the isolated dimers and directly treat 
a finite-size cluster that has not previously been reported.  
The second is to clarify the behavior of the $S=1$ case for the first time. 
Our numerical diagonalizations will provide us with a deeper understanding 
of the target system from a broader perspective. 

This paper is organized as follows. 
Section II introduces the model
and numerical method.  
Section III presents and discusses the numerical results. 
Section IV summarizes the main findings and presents concluding remarks.

\section{Model and method}

\begin{figure}
\includegraphics[width=0.85\linewidth,angle=0]{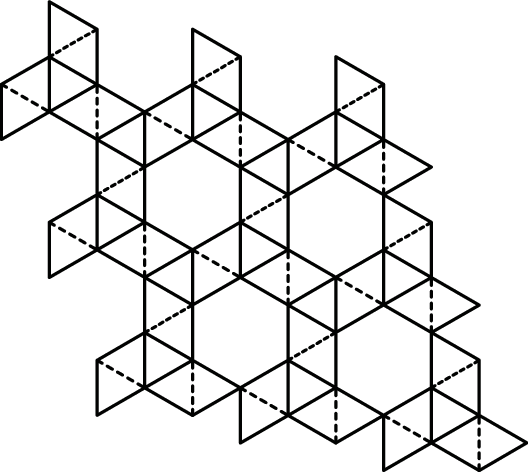}%
\caption{\label{fig1} 
Form of the target system. 
The interaction bonds $J_{\rm d}$ and $J_{\rm b}$ are represented by 
broken and solid lines, respectively. 
Solid lines form the bounce lattice. 
When the broken and solid lines are equivalent, 
the system is reduced to the maple-leaf lattice. 
}
\end{figure}

\begin{figure}
\includegraphics[width=0.85\linewidth,angle=0]{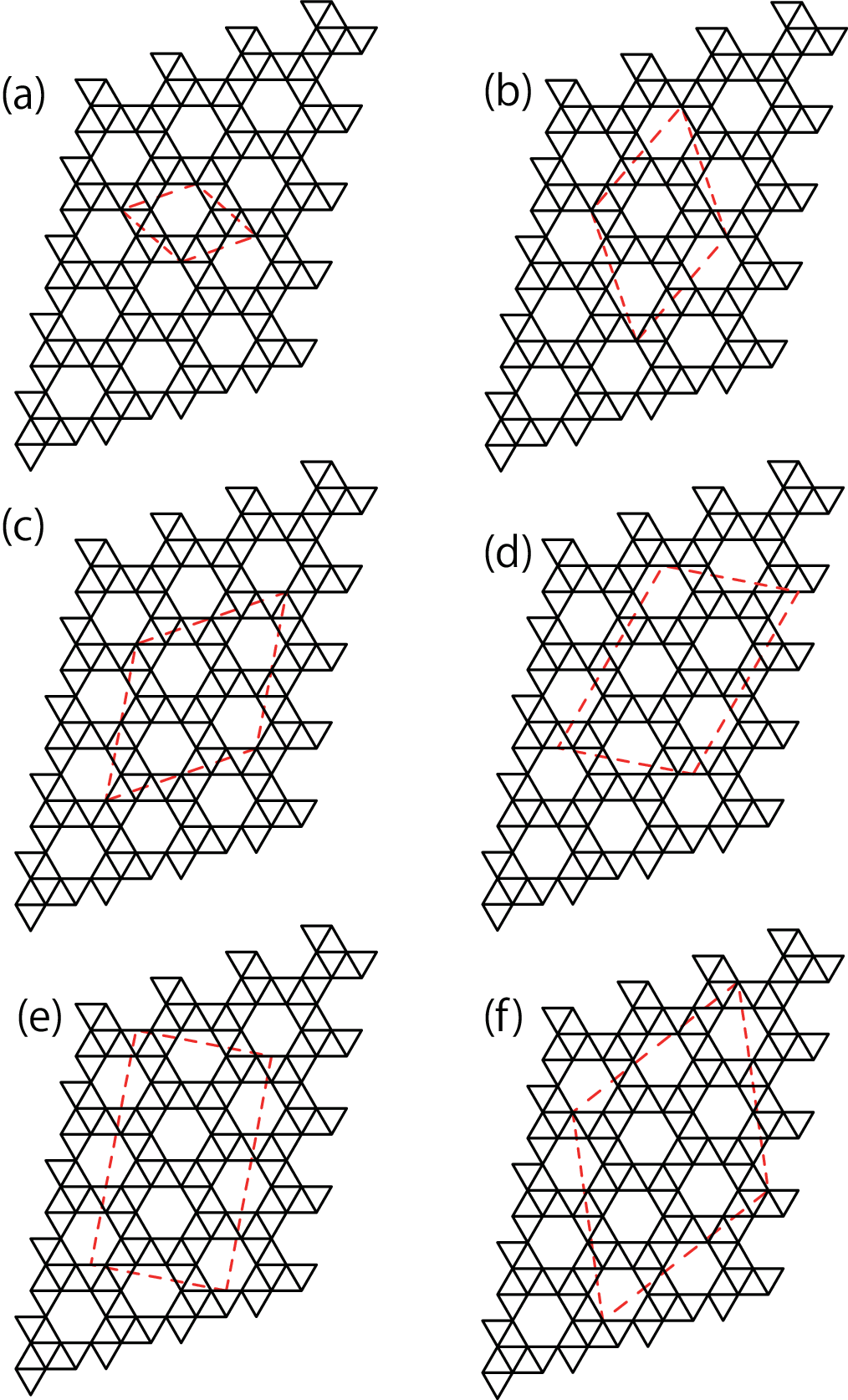}%
\caption{\label{fig2} 
Finite-size clusters under the periodic boundary condition denoted 
by broken red lines: (a) unit cell of the system ($N=6$) 
and cases (b) $N=18$,
(c) $N=24$,
(d) $N=30$,
(e) $N=36$, and 
(f) $N=42$. 
}
\end{figure}

We investigate the Heisenberg Hamiltonian, which is given by
\begin{equation}
  {\cal H} =
  \sum_{\langle i,j\rangle:~{\rm broken~bond}} J_{\rm d}
  \mbox{\boldmath $S$}_{i}\cdot \mbox{\boldmath $S$}_{j}
+ \sum_{\langle i,j\rangle:~{\rm solid~bond}} J_{\rm b}
  \mbox{\boldmath $S$}_{i}\cdot \mbox{\boldmath $S$}_{j}. 
\label{Hamiltonian}
\end{equation}
Here, $\mbox{\boldmath $S$}_{i}$ denotes the spin-$S$ spin operator 
at site $i$. 
We focus on the cases $S=1/2$ and $S=1$ for isotropic interaction 
in spin space. 
Fig.~\ref{fig1} illustrates the target system, 
whose vertices are characterized by spin sites.  
The first term of Eq.~(\ref{Hamiltonian}) represents 
dimer interactions, which are illustrated by broken bonds. 
The second term represents interactions 
that form the bounce lattice, which is illustrated by solid bonds. 
The system reduces to the maple-leaf lattice 
when the broken and solid bonds are equivalent. 
The number of spin sites is represented by $N$. 
Note that $N/6$ is an integer because 
a single unit cell of the treated system, as illustrated
in Fig.~\ref{fig2}(a), has six spin sites regardless of whether 
the broken and solid bonds are equivalent. 
We limit the study to the cases of $J_{\rm d}>0$ and $J_{\rm b}>0$ 
when these two interactions between two spins are antiferromagnetc. 
We measure energies in units of either $J_{\rm d}$ or $J_{\rm b}$. 
The suitable one depends on the circumstances. 
For small $J_{\rm d}$, a dimensionless parameter $J_{\rm d}/J_{\rm b}$ 
is used on one hand 
while on the other hand, 
another dimensionless parameter $J_{\rm b}/J_{\rm d}$ is used 
for small $J_{\rm b}$.  
When $J_{\rm b}=0$, the system is an assembly of isolated dimerized-spin models.
When $J_{\rm d}=0$, the system is the bounce-lattice antiferromagnet. 
The periodic boundary condition is employed when finite-size clusters 
with a system size $N$ are treated. 
For $S=1/2$, we treat $N=18$, 24, 30, 36, and 42, which 
are depicted in Fig.~\ref{fig2}(b)-Fig.~\ref{fig2}(f), respectively.
For $S=1$, we treat $N=18$ and 24. 
Because these are actually small system sizes, 
they are not sufficient to capture a complete understanding 
of the target system. 
It is important to make efforts to maximize the system sizes 
as large as possible 
because the number of feasible methods that can be applied 
is quite limited when the system includes frustrations. 
Such newly obtained results would determine whether 
our understanding from previous investigations would be confirmed 
or should be updated. 
The cases of $N=18$, 24, and 42 form 
a rhombus with an inner angle of $\pi/3$,  
in which the tilting angles are different from each other. 
The equivalence of three directions is maintained 
in the present two-dimensional system, which includes  
threefold rotational symmetry. 
Therefore, the two-dimensionality of the system 
is appropriately taken into account. 
In the cases of $N=30$ and 36, 
clusters cannot form the same shape. 
So, the equivalence of the three directions is not maintained. 

A marked behavior of this system is that 
the ground state is the direct product of singlet dimers 
located at $J_{\rm d}$ bonds when $J_{\rm b}$ is small 
irrespective of whether it is vanishing or nonzero. 
This behavior is independent of $S$. 
When $[j_{1},j_{2}]$ denotes 
the spin singlet state at a pair of sites $j_{1}$ and $j_{2}$,  
$[j_{1},j_{2}]$ for $S=1/2$ is expressed by
\begin{eqnarray}
  \frac{1}{\sqrt{2}}[
& &|S=\frac{1}{2},m=\frac{1}{2}\rangle_{j_{1}}\otimes|S=\frac{1}{2},m=-\frac{1}{2}\rangle_{j_{2}} 
\nonumber\\
&-&|S=\frac{1}{2},m=-\frac{1}{2}\rangle_{j_{1}}\otimes|S=\frac{1}{2},m=\frac{1}{2}\rangle_{j_{2}}].
\end{eqnarray}
For $S=1$, $[j_{1},j_{2}]$ is expressed by 
\begin{eqnarray}
  \frac{1}{\sqrt{3}}[
& & |S=1,m=1\rangle_{j_{1}}\otimes|S=1,m=-1\rangle_{j_{2}}
\nonumber \\
&-&|S=1,m=0\rangle_{j_{1}}\otimes|S=1,m=0\rangle_{j_{2}}
\nonumber \\
&+&|S=1,m=-1\rangle_{j_{1}}\otimes|S=1,m=1\rangle_{j_{2}}    
]. 
\end{eqnarray}
If we construct the exact-dimer state
$|\Phi_{\rm ED}\rangle = \prod_{(j_{1},j_{2})}[j_{1},j_{2}]$, 
it becomes an eigenstate of ${\cal H}$. 
We then obtain 
${\cal H}|\Phi_{\rm ED}\rangle=E_{\rm ED}|\Phi_{\rm ED}\rangle$, where 
\begin{equation}
E_{\rm ED} = - \frac{3}{8} J_{\rm d} N ,
\label{ene_dimer_s05}
\end{equation}
for $S=1/2$ and 
\begin{equation}
E_{\rm ED} = - J_{\rm d} N ,
\label{ene_dimer_s1}
\end{equation}
for $S=1$. 
Note that the ratio $J_{\rm b}/J_{\rm d}$ affects  
whether the eigenstate $|\Phi_{\rm ED}\rangle$ 
is the ground state of a system including $J_{\rm d}$ and $J_{\rm b}$. 
In this study, 
the edge of the region of $J_{\rm b}/J_{\rm d}$ 
in which this eigenstate is the ground state 
is examined numerically for both $S=1/2$ and $S=1$. 

We carry out numerical diagonalization by using 
the Lanczos algorithm\cite{Lanczos} 
in the subspace characterized by $M(=\sum_{j} S_{j}^{z})$ 
to obtain the lowest energies of ${\cal H}$. 
We take the $z$-axis as the quantized axis of each spin. 
One important advantage of numerical diagonalization 
is that it treats quantum effects without any approximations. 
This method yields precise results 
regardless of whether frustration is present, which allows 
reliable information about the system to be obtained. 
Because we focus on the low-energy properties 
of the target system, we have carried out diagonalization 
to obtain the lowest eigenenergy $E(N,M)$ for a given $N$ and $M$. 
The ground-state energy $E_{\rm g}$ of the system is given by $E(N,M=0)$ 
and the spin excitation gap $\Delta$ is given by
\begin{equation}
\Delta = E(N,M=1) - E(N,M=0) ,
\label{spin_gap}
\end{equation}
for each $N$.
The Lanczos diagonalization was carried out using an MPI-parallelized code 
that was originally developed in Ref.~24. 
The usefulness of our program was confirmed 
through large-scale parallelized calculations\cite{HNakano_kgm_gap_JPSJ2011,
HN_TSakai_kgm_1_3_JPSJ2014,HN_TSakai_kgm_S_JPSJ2015,HN_TSakai_kgm45_JPSJ2018,
HNakano_HaldaneGap_JPSJ2019,HNakano_HaldaneGap_JPSJ2022}.

\section{Results and discussion}

\subsection{The case of $S=1/2$}

\begin{table*}[tb]
\caption{
Numerical results for the Heisenberg antiferromagnets 
on the maple-leaf and bounce lattices in the cases of $S=1/2$. 
The ground-state energies, spin excitation gaps, 
spin correlation functions between $i$ and $j$ 
connected by interaction bonds are presented. 
Underlines indicate that the values are obtained by averaging  
the corresponding data. 
}
\label{table1}
\begin{tabular}{r|c|c|r|r|r}
\hline
$N$ & $-E_{\rm g}/(NJ_{\rm b})$ & $ \Delta/J_{\rm b} $ &
$\langle\mbox{\boldmath $S$}_{i}\cdot\mbox{\boldmath $S$}_{j}\rangle_{\rm di}$ &
$\langle\mbox{\boldmath $S$}_{i}\cdot\mbox{\boldmath $S$}_{j}\rangle_{\rm hex}$ &
$\langle\mbox{\boldmath $S$}_{i}\cdot\mbox{\boldmath $S$}_{j}\rangle_{\rm tri}$ 
\\
\hline
\multicolumn{6}{l}{maple-leaf lattice AF ($J_{\rm d}=J_{\rm b}$)} \\
\hline
18&0.547510493 & 0.545168730 & 0.0109232 & $-$0.3666731 & $-$0.1862990 \\
24&0.538076426 & 0.527993132 &$-$0.0086945 & $-$0.3737062 & $-$0.1600229 \\
30&0.538214887 & 0.439698232 & \underline{0.0061081} & \underline{$-$0.3650981} & \underline{$-$0.1761708} \\
36&0.538972382 & 0.400008593 & \underline{0.0086279} & \underline{$-$0.3655546} & \underline{$-$0.1777317} \\
42&0.528253583 & 0.388389034 &$-$0.0054313 & $-$0.3774549 & $-$0.1480831 \\
\hline
\multicolumn{6}{l}{bounce lattice AF ($J_{\rm d}=0$)} \\
\hline
18&0.583398505 & 0.597728751 & 0.1139687 & $-$0.4060996 & $-$0.1772989 \\
24&0.566306211 & 0.572350897 & 0.1051656 & $-$0.4025167 & $-$0.1637895 \\
30&0.571497948 & 0.478346539 & \underline{0.1110641} & \underline{$-$0.3997179} & \underline{$-$0.1717800} \\
36&0.573079165 & 0.445141194 & \underline{0.1116416} & \underline{$-$0.4007447} & \underline{$-$0.1723344} \\
42&0.558122037 & 0.497877203 & 0.1080995 & $-$0.3992790 & $-$0.1588430 \\
\hline
\end{tabular}
\end{table*}
Table~\ref{table1} summarizes the numerical data 
for antiferromagnets on the maple-leaf and bounce lattices 
in the  case of $S=1/2$,  
which includes 
the ground-state energy $E_{\rm g}$, 
the spin excitation gap $\Delta$, and 
the spin correlation functions  
$\langle\mbox{\boldmath $S$}_{i}\cdot\mbox{\boldmath $S$}_{j}\rangle_{\rm di}$, 
$\langle\mbox{\boldmath $S$}_{i}\cdot\mbox{\boldmath $S$}_{j}\rangle_{\rm hex}$, 
and 
$\langle\mbox{\boldmath $S$}_{i}\cdot\mbox{\boldmath $S$}_{j}\rangle_{\rm tri}$. 
The pair of $i$ and $j$ in 
$\langle\mbox{\boldmath $S$}_{i}\cdot\mbox{\boldmath $S$}_{j}\rangle_{\rm di}$ 
is placed on dimer bonds (i.e. the broken bonds in Fig.~\ref{fig1}). 
The pair of $i$ and $j$ in 
$\langle\mbox{\boldmath $S$}_{i}\cdot\mbox{\boldmath $S$}_{j}\rangle_{\rm hex}$ 
is placed on an edge of a local hexagon. 
The pair of $i$ and $j$ in 
$\langle\mbox{\boldmath $S$}_{i}\cdot\mbox{\boldmath $S$}_{j}\rangle_{\rm tri}$ 
is placed on one of the remaining interaction bonds.  
The spin correlation functions are obtained 
by $\langle S^{z}_{i}S^{z}_{j}\rangle$ in triplicate. 
For $N=30$ and 36, averaging is required 
because their corresponding clusters (Fig.~\ref{fig2}(d) and 
Fig.~\ref{fig2}(e)) are not in the shape of a rhombus. 
Numerical data up to $N=36$ 
obtained through our own calculations are supplemented with
results from Refs.~2, 3, and 4. 

\begin{figure}
\includegraphics[width=0.85\linewidth,angle=0]{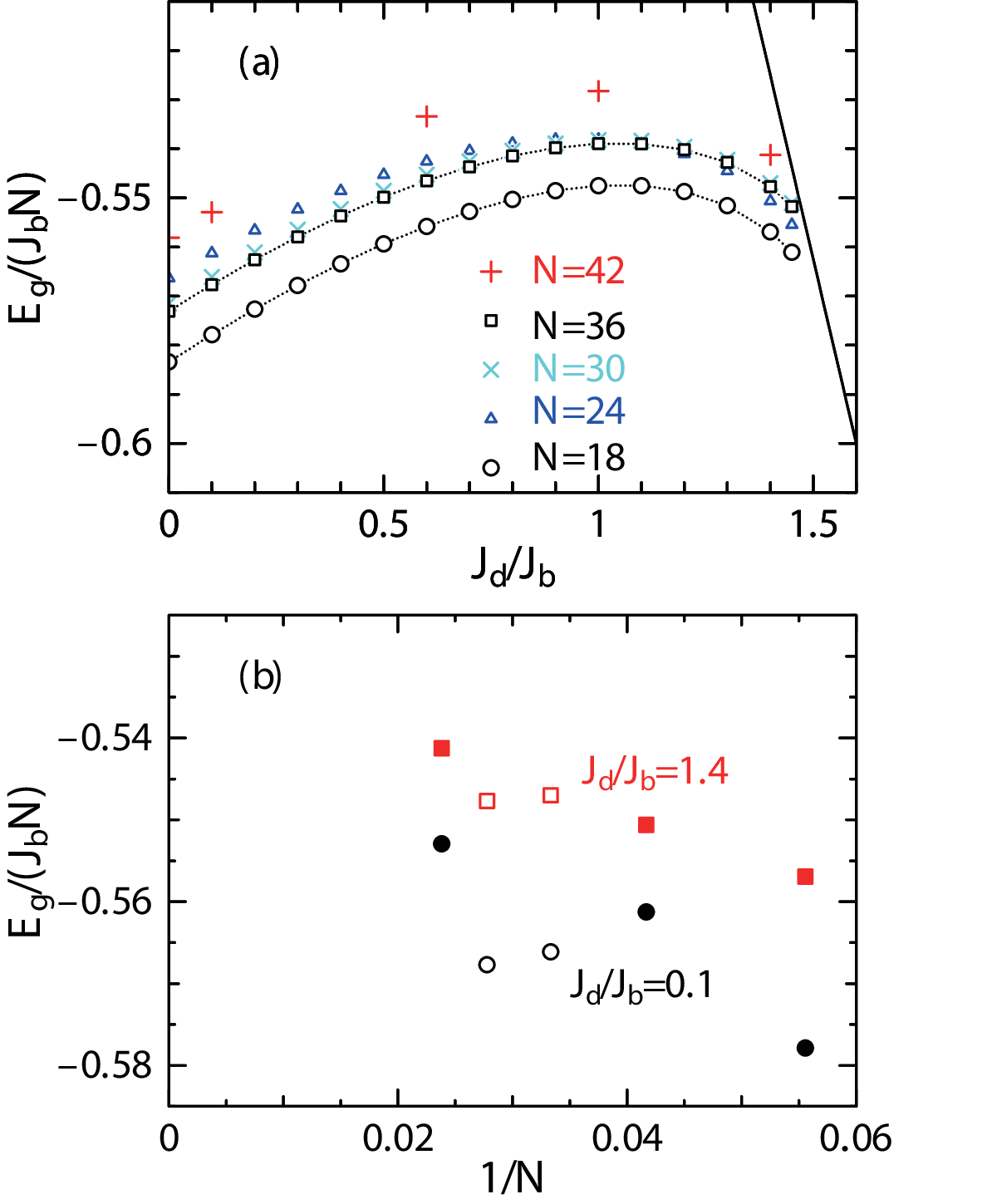}%
\caption{\label{fig3} 
Ground-state energy per site measured in units of $J_{\rm b}$ for $S=1/2$: 
(a) $N=18$ and 36 from Ref.~4 
(circles and squares linked by dotted lines) 
and additional results for $N=24$ (blue triangles),
$N=30$ (green crosses), and $N=42$ (red pluses). 
The solid line represents the energy level of the exact-dimer state. 
(b) System-size dependence for the cases of 
$J_{\rm d}/J_{\rm b}=0.1$ (black circles) and 1.4 (red squares). 
Closed and open symbols correspond 
to the cases of rhombic and nonrhombic clusters, respectively. 
}
\end{figure}

We use numerical diagonalization to examine the ground-state energy per site 
in the case of $S=1/2$. 
Fig.~\ref{fig3}(a) combines the results
for $N=18$ (Fig.~\ref{fig2}(b)) and $N=36$ (Fig.~\ref{fig2}(e)) 
reported in Ref.~4 
with our own results for $N=24$ (Fig.~\ref{fig2}(c)),
30 (Fig.~\ref{fig2}(d)), and 42 (Fig.~\ref{fig2}(f)), 
which are presented in units of $J_{\rm b}$. 
The solid line shows the energy level of the exact-dimer eigenstate. 
A ground state that is different from the exact-dimer state appears 
below $J_{\rm d}/J_{\rm b}\sim 1.5$. 
However, the system shows a nonmonotonic dependence on $N$ 
when a specific value is given as the ratio $J_{\rm d}/J_{\rm b}$. 
To clarify the $N$-dependence of the ground-state energy 
for a given $J_{\rm d}/J_{\rm b}$,  
Fig.~\ref{fig3}(b) plots the ground-state energy per site 
as a function of $1/N$. 
For the rhombic clusters, plotting three data points 
reveals weak deviation from a linear behavior. 
For $J_{\rm d}/J_{\rm b}=0.1$, 
two data points for nonrhombic clusters are significantly apart 
from the dependence of rhombic-cluster data points. 
The deviates of nonrhombic-cluster data points 
get smaller when $J_{\rm d}/J_{\rm b}$ is increased from 0.1 to 1.4. 
These features suggest that 
rhombic clusters are appropriate 
to know well the ground-state properties in the thermodynamic limit. 
In particular, one has to be careful when behaviors in the region 
near the bounce-lattice antiferromagnet are examined. 

\begin{figure}
\includegraphics[width=0.85\linewidth,angle=0]{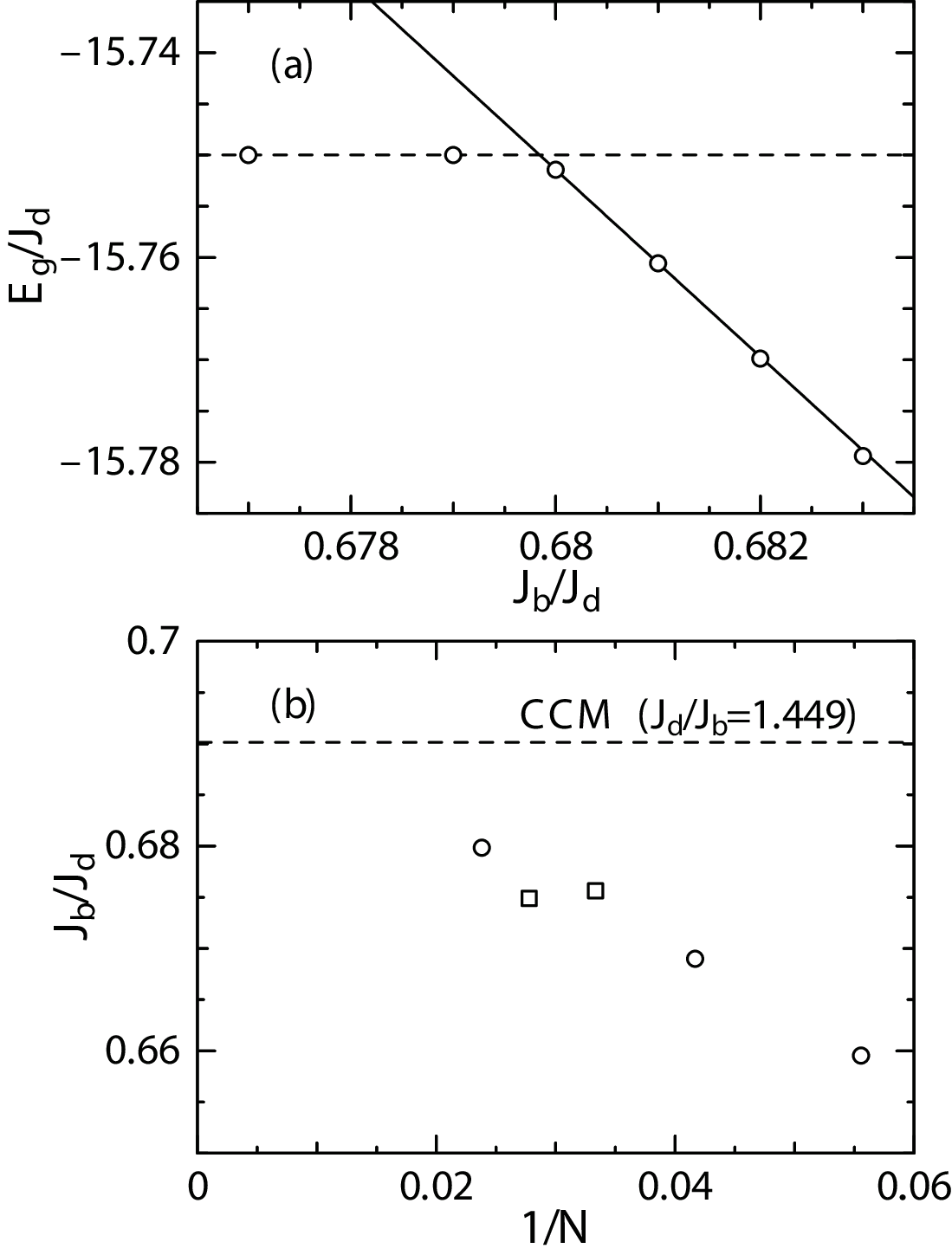}%
\caption{\label{fig4} 
(a) Edge of the exact-dimer phase in the case of $S=1/2$ 
for the $N=42$ cluster. 
Circles show the numerical diagonalization results, 
the horizontal broken line denotes the energy level of the exact-dimer state 
and the solid line is obtained by fitting data outside the exact-dimer phase. 
(b) System-size dependence of the edge 
of the exact-dimer phase for finite-size clusters. 
Circles and squares correspond to rhombic and nonrhombic clusters, 
respectively. 
The horizontal broken line shows 
the result obtained by the coupled-cluster method in Ref.~4. 
}
\end{figure}

Next, we investigate the edge of the exact-dimer phase 
in the case of $S=1/2$. 
Fig.~\ref{fig4}(a) depicts the ground-state energy for $N=42$ around the edge.  
The numerical diagonalization captures the energy level 
of the exact-dimer eigenstate (the broken horizontal line) 
for small values of $J_{\rm b}/J_{\rm d}$. 
When $J_{\rm b}/J_{\rm d}$ becomes larger than a certain value,
the ground-state energy becomes lower 
than the energy level of the exact-dimer state. 
We then draw a fitting line from two data points 
close to the energy level of the exact-dimer state. 
Other data points for even larger $J_{\rm b}/J_{\rm d}$ fall on the fitting line. 
The lower levels strongly suggest the emergence
of a new spin state that differs from the exact-dimer state 
due to the level crossing. 
The crossing point of the horizontal line and fitting line for $N=42$ at 
\begin{equation}
J_{\rm d}/J_{\rm b}=0.67984 . 
\end{equation}
Fig.~\ref{fig4}(b) shows the $N$-dependence of the crossing point 
for finite-size clusters at the edge of the exact-dimer phase 
along with result obtained by the coupled-cluster method 
(CCM)\cite{Farnell_Richter_PRB2011}. 
Our results for finite-size rhombic clusters 
obtained by numerical diagonalization come very close to the CCM result. 

\begin{figure}
\includegraphics[width=0.85\linewidth,angle=0]{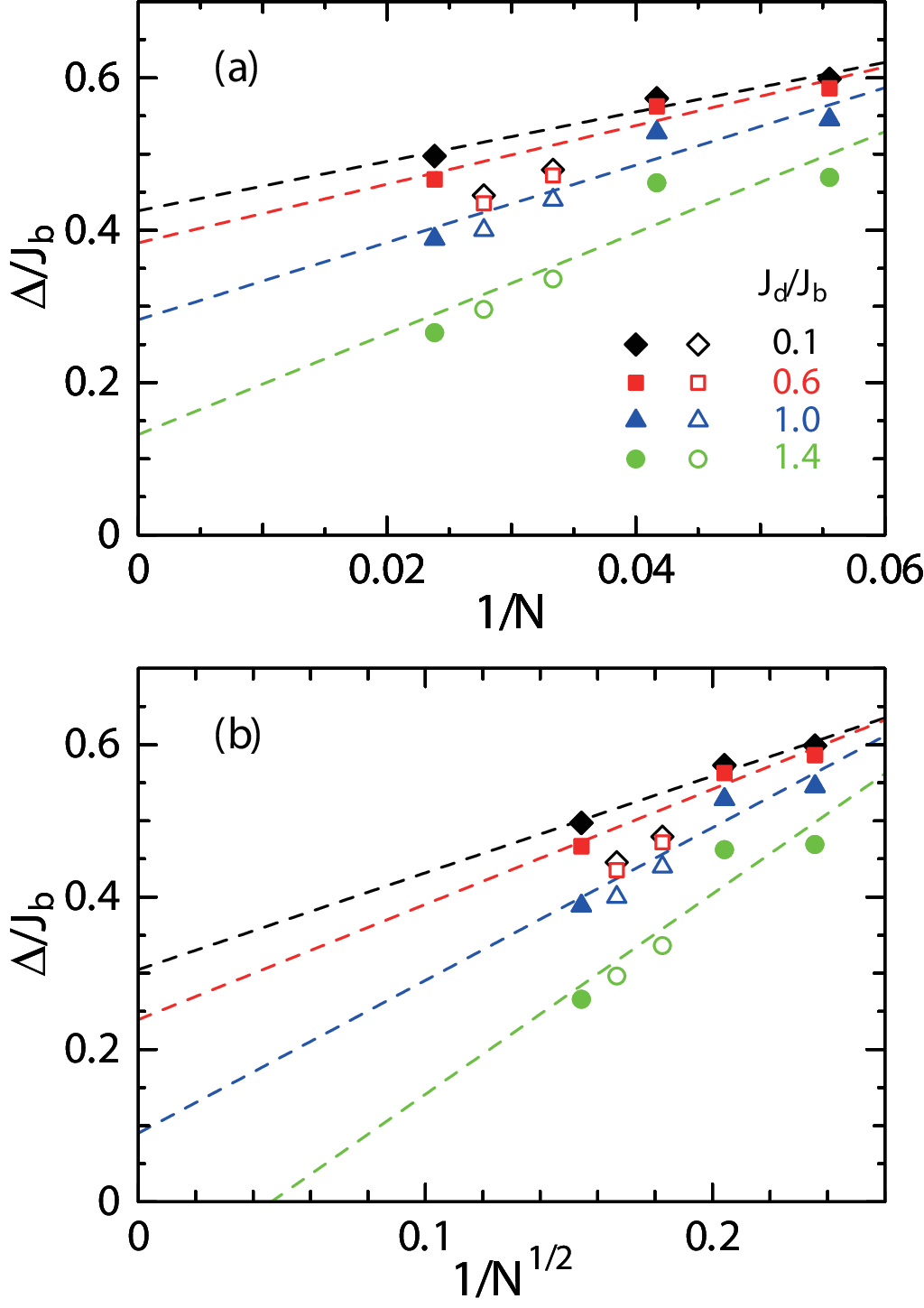}%
\caption{\label{fig5} 
Extrapolation of the spin excitation gap for $S=1/2$ 
as functions of (a) $1/N$ 
and (b) $1/\sqrt{N}$.  
The numerical-diagonalization results 
for $J_{\rm d}/J_{\rm b}=0.1$, 0.6, 1.0, and 1.4 by black diamonds, 
red squares, blue triangles, and green circles, respectively. 
Closed and open symbols correspond 
to rhombic and nonrhombic clusters, respectively. 
Broken lines show the fitting results based on data from the rhombic clusters.
}
\end{figure}

\begin{figure}
\includegraphics[width=0.85\linewidth,angle=0]{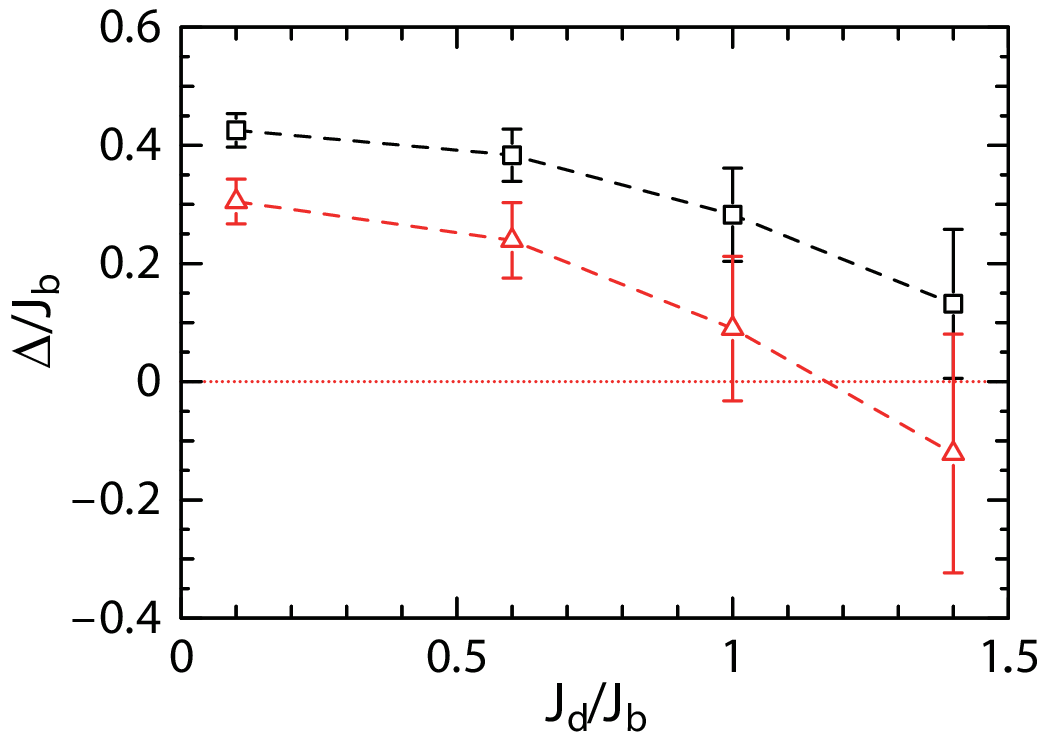}%
\caption{\label{fig6} 
Fitting error bars for extrapolation analysis of the spin excitation gap 
for $S=1/2$  
using $1/N$ (black squares) and $1/\sqrt{N}$ (red triangles) 
as the abscissa. 
}
\end{figure}

Fig.~\ref{fig5} plots the spin excitation gap in the case of $S=1/2$ 
for finite-size clusters as a function of (a) $1/N$
and (b) $1/\sqrt{N}$. 
Ref.~3  
also plotted the spin excitation gap 
as a function of $1/N$
to analyze the maple-leaf-lattice antiferromagnet. 
In Fig.~\ref{fig5}(a), 
plotting three data points for the rhombic clusters 
results in linear fitting lines for each $J_{\rm d}/J_{\rm b}$ case 
(broken lines). 
The fitting errors obtained are shown in Fig.~\ref{fig6} later.
The intercept on the ordinate becomes smaller 
with increasing $J_{\rm d}/J_{\rm b}$.  
For nonrhombic clusters, plotting two data points  
reveals notable deviations from the linear fitting line 
in the cases of $J_{\rm d}/J_{\rm b}=0.1$ and 0.6. 
In contrast, the deviations are smaller for $J_{\rm d}/J_{\rm b}=1.0$ and 1.4. 
For $J_{\rm d}/J_{\rm b}=1.0$ and 1.4, 
plotting four data points for $N\ge 24$ clusters reveals 
another linear dependence with an intercept that is smaller 
than the intercept of the fitting line. 
For $J_{\rm d}/J_{\rm b}=1.4$, plotting four data points for $N\ge 24$ clusters 
seems to result in a vanishing intercept. 
Based on these features, 
an analysis based on Fig.~\ref{fig5}(a) 
is likely to overestimate the spin excitation gap. 
However, it would still be useful for identifying general characteristics.

Figures~\ref{fig5}(a) and \ref{fig5}(b) differs in the abscissa.  
In two-dimensional systems, 
$\sqrt{N}$ corresponds to a characteristic length of the system. 
Ref.~31 
also plotted the spin excitation gap of finite-size clusters 
as a function of 
the inverse of the characteristic length of the system 
to analyze the kagome-lattice antiferromagnet. 
For rhombic finite-size clusters, plotting three data points 
results in a linear fitting line (broken lines) in Fig.~\ref{fig5}(b). 
For $J_{\rm d}/J_{\rm b}=1.4$,
the intercept of the linear fitting line is negative,  
which suggests that the spin excitation gap is absent. 
For $J_{\rm d}/J_{\rm b}=1.0$, the intercept is positive but small. 
For $J_{\rm d}/J_{\rm b}=0.1$ and 0.6, 
the intercepts are large and nonvanishing values,  
which suggests that these cases are gapped. 
Based on these features, 
an analysis based on Fig.~\ref{fig5}(b) 
is likely to underestimate the spin excitation gap. 

Fig.~\ref{fig6} depicts the intercept from the previous analysis 
on the spin excitation gap accompanied by the fitting errors. 
For $J_{\rm d}/J_{\rm b}=1.4$, 
the intercepts vanishes within their error bars 
regardless of whether the analysis is 
based on $1/N$ (Fig.~\ref{fig5}(a)) or $1/\sqrt{N}$ (Fig.~\ref{fig5}(b)). 
For $J_{\rm d}/J_{\rm b}=1.0$, 
the intercept vanishes within its error bar 
when the analysis based on $1/\sqrt{N}$ but not when it is based on $1/N$. 

These results do not conclusively determine 
whether the system is gapless or gapped. 
Hereafter, the term ``gapless'' refers to a vanishing spin gap 
in the thermodynamic limit within error bars. 
At present, they are not inconsistent with the result in Ref.~3 
to be the gapless excitation for $J_{\rm d}/J_{\rm b}=1.0$. 
For $J_{\rm d}/J_{\rm b}=0.6$ and 0.1, the intercepts are nonzero 
within their errors regardless of whether the analysis is 
based on Fig.~\ref{fig5}(a) or Fig.~\ref{fig5}(b). 
These results suggest that the system is gapped 
in the region around the bounce-lattice antiferromagnet. 
Generally speaking, if a system is gapped, 
this indicates that its ground state has no long-range order. 
Therefore, the case around the bounce-lattice antiferromagnet 
should have no long-range order in the ground state. 
This gapped consquence does not agree with Ref.~4, 
which concluded 
that a magnetic order survives in this region. 
This discrepancy must be resolved in future studies. 
Recent two studies based on the tensor-network calculations 
in Ref.~17 
and 
Ref.~19 
reported conclusions for small $J_{\rm d}/J_{\rm b}$ 
that differ from each other. 
In Ref.19 
nonzero local spin moment was reported in the region 
$0 \le J_{\rm d}/J_{\rm b} \le 1.4$. 
The behavior was in agreement with Ref.4. 
On the other hand, 
Ref.~17 
showed vanishing staggered magnetization for small $J_{\rm d}/J_{\rm b}$ 
while nonzero and significant staggered magnetization appears 
in the region around $J_{\rm d}/J_{\rm b}\sim 1.3$. 

\noindent
Our results also suggest 
that there appear the gapless cases 
between this gapped region for small $J_{\rm d}/J_{\rm b}$ 
and the region of the exact-dimer state. 
This behavior from 
Ref.~17 
is consistent with the present numerical-diagonalization study. 
The present results of the gapless nature 
do not contradict those of Ref.~4, 
concerning 
the ground state of the maple-leaf-lattice antiferromagnet. 

\subsection{The case of $S=1$}

\begin{table*}[tb]
\caption{
Numerical results for the Heisenberg antiferromagnets 
on the maple-leaf and bounce lattices in the case of $S=1$. 
The ground-state energies, spin excitation gaps, 
spin correlation functions between $i$ and $j$ 
connected by interaction bonds are presented. 
}
\label{table2}
\begin{tabular}{r|c|c|r|r|r}
\hline
$N$ & $-E_{\rm g}/(NJ_{\rm b})$ & $ \Delta/J_{\rm b} $ &
$\langle\mbox{\boldmath $S$}_{i}\cdot\mbox{\boldmath $S$}_{j}\rangle_{\rm di}$ &
$\langle\mbox{\boldmath $S$}_{i}\cdot\mbox{\boldmath $S$}_{j}\rangle_{\rm hex}$ &
$\langle\mbox{\boldmath $S$}_{i}\cdot\mbox{\boldmath $S$}_{j}\rangle_{\rm tri}$ 
\\
\hline
\multicolumn{6}{l}{maple-leaf lattice AF ($J_{\rm d}=J_{\rm b}$)} \\
\hline
18&1.764156308 & 0.483132650 & $-$0.0027243 & $-$1.1345011 & $-$0.6282931 \\
24&1.705276761 & 0.471956065 & $-$0.0249702 & $-$1.1815384 & $-$0.5112533 \\
\hline
\multicolumn{6}{l}{bounce lattice AF ($J_{\rm d}=0$)} \\
\hline
18&1.904791072 & 0.556416135 & 0.4781426 & $-$1.2870948 & $-$0.6176962 \\
24&1.825846125 & 0.545393164 & 0.4428176 & $-$1.2658120 & $-$0.5600342 \\
\hline
\end{tabular}
\end{table*}
Table~\ref{table2} summarizes the numerical data for 
antiferromagnets on the maple-leaf and bounce lattices 
for the case of $S=1$,  
which includes 
$E_{\rm g}$, $\Delta$, 
$\langle\mbox{\boldmath $S$}_{i}\cdot\mbox{\boldmath $S$}_{j}\rangle_{\rm di}$, 
$\langle\mbox{\boldmath $S$}_{i}\cdot\mbox{\boldmath $S$}_{j}\rangle_{\rm hex}$, 
and 
$\langle\mbox{\boldmath $S$}_{i}\cdot\mbox{\boldmath $S$}_{j}\rangle_{\rm tri}$ 
as in the $S=1/2$ case. 

\begin{figure}
\includegraphics[width=0.85\linewidth,angle=0]{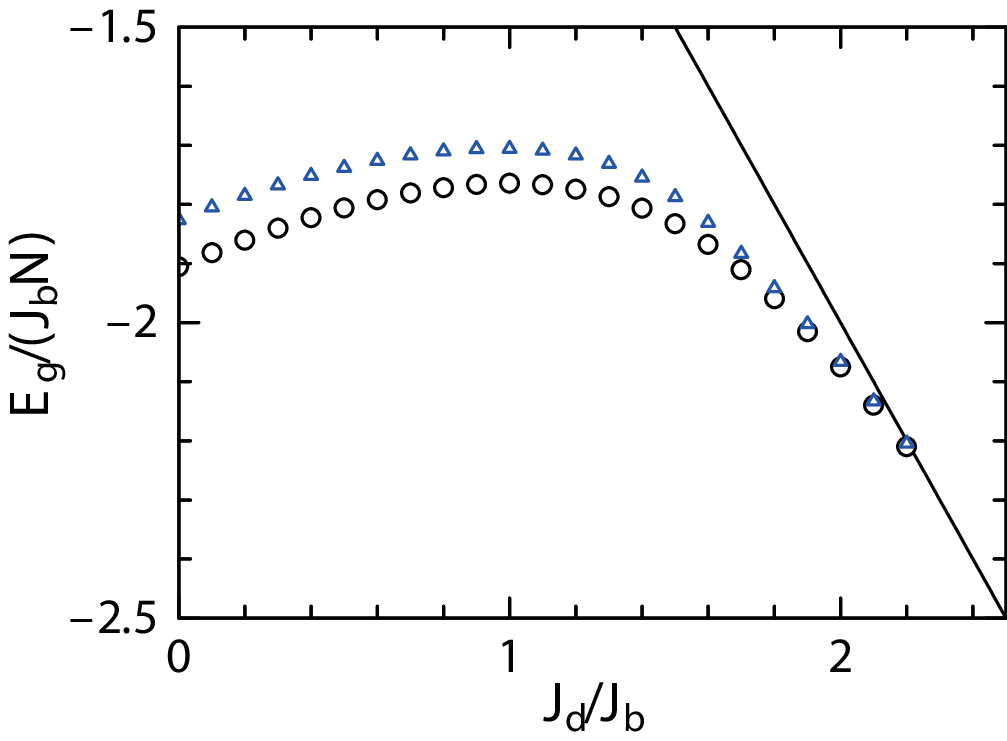}%
\caption{\label{fig7} 
Ground-state energy per site measured in units of $J_{\rm b}$ for $S=1$:  
$N=18$ (black circles) and $N=24$ (blue triangles). 
The solid line represents the energy level of the exact-dimer state. 
}
\end{figure}

Fig.~\ref{fig7} shows the ground-state energy per site for $N=18$ and 24 
as a function of $J_{\rm d}/J_{\rm b}$ in units of $J_{\rm b}$ along with 
the energy level of the exact-dimer eigenstate. 
This figure is the $S=1$ version of Fig.~\ref{fig3}(a) 
for $S=1/2$. 
The ground-state energy gradually increases 
with increasing $J_{\rm d}/J_{\rm b}$ up to $J_{\rm d}/J_{\rm b}\sim 1$. 
As $J_{\rm d}/J_{\rm b}$ is further increased,  
the ground-state energy gradually decreases 
to meet the energy level of the exact-dimer state 
at $J_{\rm d}/J_{\rm b}\sim 2.2$. 

\begin{figure}
\includegraphics[width=0.85\linewidth,angle=0]{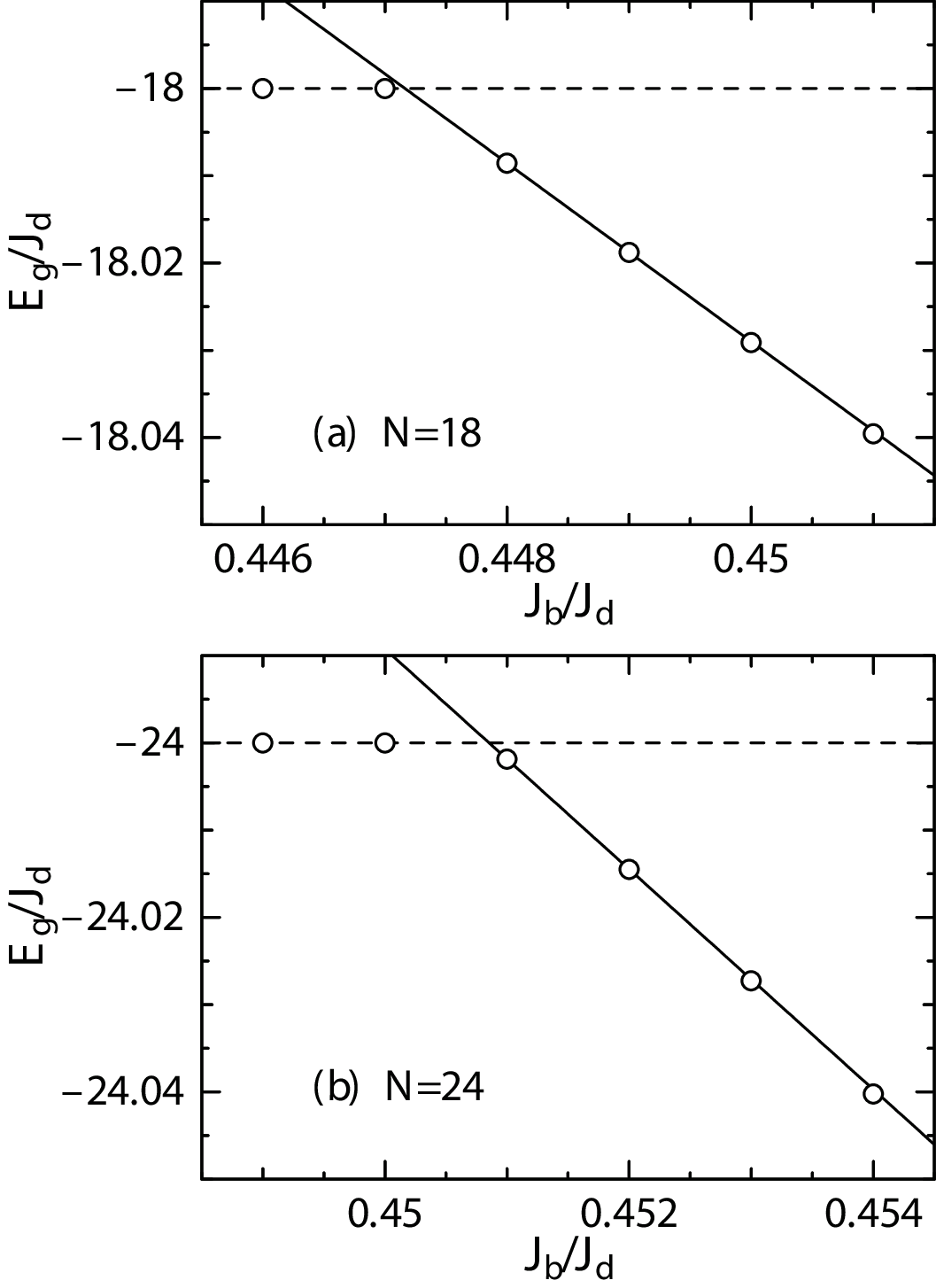}%
\caption{\label{fig8} 
Ground-state energy around the edge of the exact-dimer phase for $S=1$: 
(a) $N=18$ and (b) $N=24$. 
Circles correspond to results from the numerical diagonalization. 
The horizontal broken line denotes the energy level of the exact-dimer state. 
The solid line is obtained by fitting data outside the exact-dimer phase. 
}
\end{figure}

To capture the behavior at the edge of the exact-dimer phase 
in detail, 
Fig.~\ref{fig8} depicts the ground-state energy in units of $J_{\rm d}$ 
as a function of $J_{\rm b}/J_{\rm d}$ for (a) $N=18$ and (b) $N=24$.  
The numerical diagonalization successfully captures the energy level 
of the exact-dimer eigenstate (broken horizontal line)  
for small values of $J_{\rm b}/J_{\rm d}$. 
When $J_{\rm b}/J_{\rm d}$ becomes larger than a certain value, 
the ground-state energy becomes lower than the energy level 
of the exact-dimer state. 
We draw a fitting line determined from two data points 
close to the energy level of the exact-dimer state. 
The crossing point between the horizontal line and fitting line 
suggests that 
\begin{equation}
J_{\rm b}/J_{\rm d}=0.44716 , 
\label{rc1s1n18}
\end{equation}
for $N=18$ and
\begin{equation}
J_{\rm b}/J_{\rm d}=0.45085 , 
\label{rc1s1n24}
\end{equation}
for $N=24$.  
The difference between Eq.~(\ref{rc1s1n18}) and Eq.~(\ref{rc1s1n24}) 
is quite small, 
which suggests that the edge of the exact-dimer phase 
for the $S=1$ system is
\begin{equation}
J_{\rm b}/J_{\rm d}\sim 0.45.  
\label{rc1s1final}
\end{equation}
This is consistent with the value of $J_{\rm d}/J_{\rm b}\sim 2.2$ 
in Fig.~\ref{fig7}. 
Note that this ratio is larger than the results for $S=1/2$ (Fig.~\ref{fig4}).

\begin{figure}
\includegraphics[width=0.85\linewidth,angle=0]{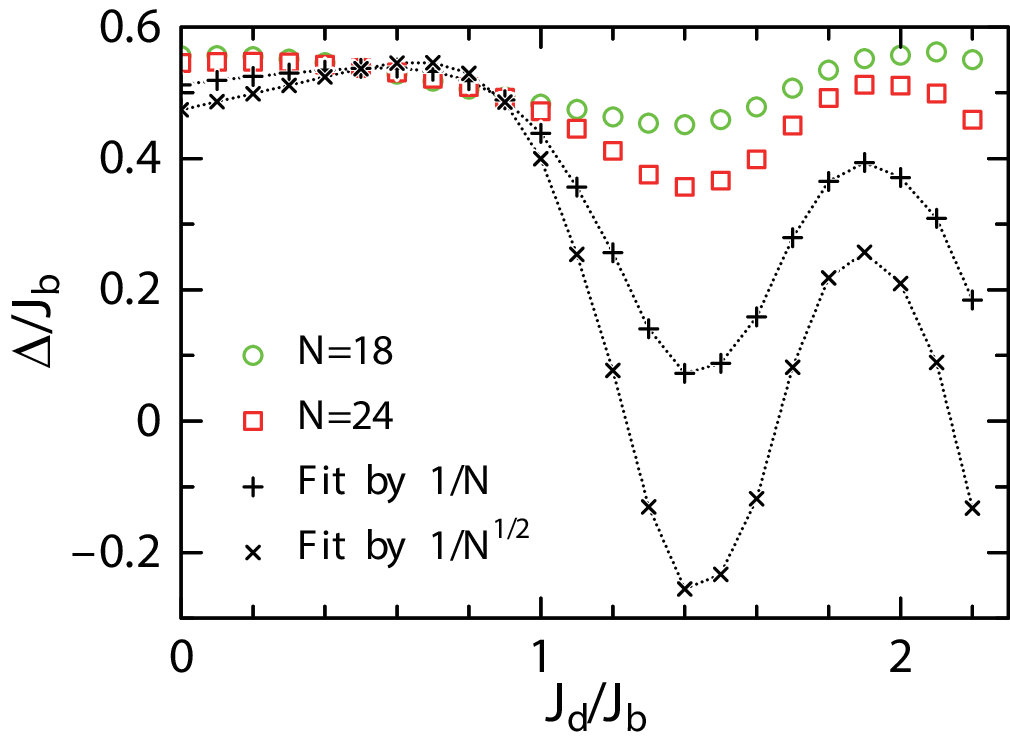}%
\caption{\label{fig9} 
Spin excitation gap for $S=1$: $N=18$ (green circles) and 
$N=24$ (red squares). 
Raw data are analyzed by plotting the spin excitation gap as functions of 
$1/N$ or $1/\sqrt{N}$; results of their extrapolation are 
represented by pluses or crosses, respectively. 
}
\end{figure}

Fig.~\ref{fig9} depicts the spin excitation gap for $S=1$. 
For finite-size clusters of $N=18$ and 24, 
the raw data do not increase monotonically with  $J_{\rm d}/J_{\rm b}$ 
up to $J_{\rm d}/J_{\rm b}\sim 2.2$. 
To examine the behavior of the spin excitation gap at large $N$, 
we carry out the same analysis as shown in Fig.~\ref{fig5}. 
The intercepts are obtained without errors 
because only two samples are obtained for each $J_{\rm d}/J_{\rm b}$ 
in the case of $S=1$. 
Up to $J_{\rm d}/J_{\rm b}\sim 1$, 
the system is gapped regardless of whether the analysis is 
based on $1/N$ or $1/\sqrt{N}$. 
This gapped nature strongly suggests that 
the ground state around the bounce-lattice antiferromagnet 
has no long-range order when $S=1$. 
On the other hand, 
the intercept is quite small at $J_{\rm d}/J_{\rm b}\sim 1.4$. 
If we remember that analyses based on $1/N$ 
are likely to result in overestimate 
and that the analyses based on $1/\sqrt{N}$ 
are likely to be underestimate, 
these results suggest that 
the system is gapless when $J_{\rm d}/J_{\rm b}\sim 1.4$. 
This suggests that the behavior of the ground state in the case of $S=1$ 
is the same as that in the case of $S=1/2$ and classical case 
reported in Ref.~3. 
At present, it is unclear whether these cases show 
a region with a nonzero width or a single point 
with respect to $J_{\rm d}/J_{\rm b}$. 
In addition, the cases of $S=1/2$ and $S=1$ share gapless behavior 
at $J_{\rm d}/J_{\rm b}\sim 1.4$. 
Further increasing $J_{\rm d}/J_{\rm b}$ results in a positive intercept again 
regardless of whether the analysis is based on $1/N$ or $1/\sqrt{N}$. 
This suggests that 
a gapped region appears below the boundary of the exact-dimer phase.  
The gapless behavior at $J_{\rm d}/J_{\rm b}\sim 1.4$ 
suggests that 
this novel gapped region differs from the cases 
of the gapped bounce-lattice antiferromagnet 
even though the long-range order also disappears owing to the presence 
of the gap, which is similar to the gapped case at $J_{\rm d}/J_{\rm b}\sim 0$. 
In the new gapped region, the intercepts decrease again 
when $J_{\rm d}/J_{\rm b}$ approaches the boundary of the exact-dimer phase. 

\section{Summary and remarks}
We have used numerical diagonalization to 
study the Heisenberg antiferromagnet on the two-dimensional lattice 
from the bounce lattice through the maple-leaf lattice 
to the isolated dimers 
in the cases of $S=1/2$ and $S=1$. 
We have primarily examined the spin excitation gap of the target system. 
Our results have suggested the presence of a gapped region 
including the case of the bounce lattice 
and gapless behavior has been observed at $J_{\rm d}/J_{\rm b}\sim 1.4$. 
This behavior does not change between the cases of $S=1/2$ and $S=1$. 
For $S=1$, 
another gapped region appears adjacent to the exact-dimer phase. 

Reference~32 
reported that Na$_{2}$Mn$_{3}$O$_{7}$ reveals 
an $S=3/2$ antiferromagnet on the lattice which is similar 
to the maple-leaf lattice. 
In this study, this material was examined experimentally  
in terms of neutron diffraction, dc magnetization, and heat capacity 
and a distortion of the maple-leaf lattice was observed 
around the $S=3/2$ magnetic ion Mn$^{4+}$. 
This study did not observe 
gapped behavior in the magnetization curve at 2K. 
This experimental behavior is not inconsistent 
with numerical result indicating gapless behavior 
around the maple-leaf lattice for $S=1/2$ and 1. 
Further theoretical studies on the $S=3/2$ case 
are required to clarify the relationship 
between experimental observations and theoretical calculations. 

The present system should be further investigated from various viewpoints. 
Such studies would deepen understanding on the effects 
of frustration in various magnetic materials. 

\begin{acknowledgments}

This work was partly supported by JSPS KAKENHI Grant Numbers 
23K11125 and 25K07229. 
In this research, we used the computational resources of the supercomputer 
Fugaku provided by RIKEN through the HPCI System Research projects 
(Project IDs: hp230114, hp230532, hp230537, and hp250164). 
Some of the computations were performed using the facilities 
of the Institute for Solid State Physics, 
The University of Tokyo and Supercomputing Division, 
Information Technology Center, The University of Tokyo. 
This research partly used computational resources of Pegasus provided
by Multidisciplinary Cooperative Research Program in Center for
Computational Sciences, University of Tsukuba.
\end{acknowledgments}


\begin{thebibliography}{99}
\bibitem{Betts}
\label{Betts}
D.~D.~Betts, 
Proc.~N.~S.~Inst.~Sci. {\bf 40}, 95 (1995). 
\bibitem{Schulenberg_Richter_APPA2000}
\label{Schulenberg_Richter_APPA2000}
J.~Schulenberg, J.~Richter, and D.~D.~Betts, 
Acta~Phys.~Pol.~A {\bf 97}, 971 (2000). 
\bibitem{Shumalfuss_Richter_PRB2002}
\label{Shumalfuss_Richter_PRB2002}
D.~Shumalfu\ss, P.~Tomczak, J.~Schulenberg, J.~Richter, 
Phys.~Rev.~B {\bf 65}, 224405 (2002).
\bibitem{Farnell_Richter_PRB2011}
\label{Farnell_Richter_PRB2011}
D.~J.~J.~Farnell, R.~Darradi, R.~Schmidt, and J.~Richter,
Phys.~Rev.~B {\bf 84}, 104406 (2011).
\bibitem{Farnell_Richter_PRB2014}
\label{Farnell_Richter_PRB2014}
D.~J.~J.~Farnell, O.~G\"{o}tze, J.~Richter, R.~F.~Bishop,
and P.~H.~Y.~Li, 
Phys.~Rev.~B {\bf 89}, 184407 (2014).
\bibitem{Farnell_PRB2018}
\label{Farnell_PRB2018}
D.~J.~J.~Farnell, O.~G\"{o}tze, J.~Schulenberg, R.~Zinke, R.~F.~Bishop,
and P.~H.~Y.~Li, 
Phys.~Rev.~B {\bf 98}, 224402 (2018).
\bibitem{Ghosh_PRBL2022}
P.~Ghosh, T.~M\"{u}ller, and R.~Thomale, 
Phys.~Rev.~B {\bf 105}, L180412 (2022).
\bibitem{Ghosh_PRB2023}
\label{Ghosh_PRB2023}
P.~Ghosh, J.~Seufert, T.~M\"{u}ller, F.~Mila, R.~Thomale, 
Phys.~Rev.~B {\bf 108}, L060406 (2023).
\bibitem{Gresista_PRB2023}
\label{Gresista_PRB2023}
L.~Gresista, G.~Hickey, S.~Trebst, Y.~Iqbal, 
Phys.~Rev.~B {\bf 108}, L241116 (2023).
\bibitem{Ghosh_JPCM2024}
\label{Ghosh_JPCM2024}
P.~Ghosh,
J. Phys.: Condens. Matter {\bf 36}, 455803 (2024).
\bibitem{Gembe_PRB2024}
\label{Gembe_PRB2024}
M.~Gemb\'{e}, L.~Gresista, H.-J.~Schmidt, C.~Hickey, Y.~Iqbal, 
S.~Trebst, 
Phys.~Rev.~B {\bf 110}, 085151 (2024).
\bibitem{Ghosh_PRB2024}
\label{Ghosh_PRB2024}
P.~Ghosh, T.~M\"{u}ller, Y.~Iqbal, R.~Thomale, H.~O.~Jeschke, 
Phys.~Rev.~B {\bf 110}, 094406 (2024).
\bibitem{Beck_PRB2024}
\label{Beck_PRB2024}
J.~Beck, J.~Bodky, J.~Motruk, T.~M\"{u}ller, R.~Thomale, P.~Ghosh, 
Phys.~Rev.~B {\bf 109}, 184422 (2024).
\bibitem{Maldonado_Mg3TeO6_PRB2025}
\label{Maldonado_Mg3TeO6_PRB2025}
C.~Aguilar-Maldonado, R.~Feyerherm, K.~Proke\v{s}, L.~Keller,
and B.~Lake, 
Phys.~Rev.~B {\bf 111}, 094439 (2025).
\bibitem{Ghosh_PRB2025}
\label{Ghosh_PRB2025}
P.~Ghosh, 
Phys.~Rev.~B {\bf 111}, 224431 (2025).
\bibitem{Hutak_PRB2025}
\label{Hutak_PRB2025}
T.~Hutak,
Phys.~Rev.~B {\bf 112}, 104405 (2025).
\bibitem{Schmoll_arXiv2407}
\label{Schmoll_arXiv2407}
P.~Schmoll, J.~Naumann, E.~L.~Weerda, J.~Eisert, and Y.~Iqbal, 
arXiv:2407.07145. 
\bibitem{Gresista_arXiv2025}
\label{Gresista_arXiv2025}
L.~Gresista, D.~Kiese, S.~Trebst, and Y.~Iqbal, 
arXiv:2511.21598. 
\bibitem{Nyckees_arXiv2512}
\label{Nyckees_arXiv2512}
S.~Nyckees, P.~Ghosh, and F.~Mila, 
arXiv:2512.20466.
%
%
\bibitem{ShastrySutherland1981}
\label{ShastrySutherland1981}
B.~S.~Shastry and B.~Sutherland, Physica B+C \textbf{108}, 1069 (1981). 
\bibitem{Kageyama_PRL1999}
\label{Kageyama_PRL1999}
H.~Kageyama, K.~Yoshimura, R.~Stern, N.~V.~Mushnikov,
K.~Onizuka, M.~Kato, K.~Kosuge, C.~P.~Slichter, T.~Goto,
and Y.~Ueda, Phys.~Rev.~Lett. {\bf 82}, 3168 (1999).
\bibitem{HNakano_TSakai_JPCM2024}
\label{HNakano_TSakai_JPCM2024}
H.~Nakano and T.~Sakai, 
J.~Phys.:~Condens.~Matter {\bf 36}, 455805 (2024). 
%
\bibitem{Haraguchi_Cu6IO3OH10CL_PRB2021}
\label{Haraguchi_Cu6IO3OH10CL_PRB2021}
Y.~Haraguchi, A.~Matsuo, K.~Kindo, Z.~Hiroi, 
Phys.~Rev.~B {\bf 104}, 174439 (2021). 
\bibitem{He_PRB2024}
\label{He_PRB2024}
A.~Lei~He, X.~H.~Yan, L.~Qi, Y.~Liu, and Y.~Han, 
Phys.~Rev.~B {\bf 109}, 075118 (2024).
%
\bibitem{Lanczos}
\label{Lanczos}
C. Lanczos,
J.~Res.~Natl.~Bur.~Stand., {\bf 45}, 255 (1950).
%
\bibitem{HNakano_HaldaneGap_JPSJ2009}
\label{HNakano_HaldaneGap_JPSJ2009}
H.~Nakano and A.~Terai,
J.~Phys.~Soc.~Jpn. {\bf 78}, 014003 (2009).
%
\bibitem{HNakano_kgm_gap_JPSJ2011}
\label{HNakano_kgm_gap_JPSJ2011}
H.~Nakano and T.~Sakai, 
J.~Phys.~Soc.~Jpn. \textbf{80}, 053704 (2011).
\bibitem{HN_TSakai_kgm_1_3_JPSJ2014}
\label{HN_TSakai_kgm_1_3_JPSJ2014}
H.~Nakano and T.~Sakai, 
J.~Phys.~Soc.~Jpn. \textbf{83}, 104710 (2014).
\bibitem{HN_TSakai_kgm_S_JPSJ2015}
\label{HN_TSakai_kgm_S_JPSJ2015}
H.~Nakano and T.~Sakai, 
J.~Phys.~Soc.~Jpn. \textbf{84}, 063705 (2015).
\bibitem{HN_TSakai_kgm45_JPSJ2018}
\label{HN_TSakai_kgm45_JPSJ2018}
H.~Nakano and T.~Sakai, 
J.~Phys.~Soc.~Jpn. \textbf{87}, 063706 (2018). 
\bibitem{HNakano_HaldaneGap_JPSJ2019}
\label{HNakano_HaldaneGap_JPSJ2019}
H.~Nakano, N.~Todoroki, and T.~Sakai,
J.~Phys.~Soc.~Jpn. {\bf 88}, 114702 (2019).
\bibitem{HNakano_HaldaneGap_JPSJ2022}
\label{HNakano_HaldaneGap_JPSJ2022}
H.~Nakano, H.~Tadano, N.~Todoroki, and T.~Sakai,
J.~Phys.~Soc.~Jpn. {\bf 91}, 074701 (2022).
%
\bibitem{Iqbal_kagome_PRB2023}
Y.~Iqbal, D.~Poilblanc, and F.~Becca, 
Phys.~Rev. B {\bf 89}, 020407(R) (2014). 
  %
\bibitem{Saha_Na2Mn3O7_PRB2023}
\label{Saha_Na2Mn3O7_PRB2023}
B.~Saha, A.~K.~Bera, S.~M.~Yusuf, and A.~Hoser, 
Phys.~Rev.~B {\bf 107}, 064419 (2023).
\end{thebibliography}

\end{document}